\documentclass[12pt]{article}
\usepackage{amsmath,amssymb,bm,bbm,amsthm} 
\usepackage[unicode=true,colorlinks=true,linkcolor=black, urlcolor=blue, citecolor = black,breaklinks]{hyperref}
\newcommand{\ket}[1]{| #1 \rangle}

\newcommand{\brkt}[2]{\langle #1 | #2 \rangle} 
\newcommand{\Eq}[1]{Eq.~(\ref{#1})}
\newcommand{\Sec}[1]{Section~\ref{Sec:#1}}
\newcommand{\mi}{i}%{\imath}
\newcommand{\me}{e} 
\newcommand{\cc}{\Bar}
\newcommand{\ul}[1]{\underline{\smash{#1}\vphantom{0}}} 
\newcommand{\SAlt}{\tilde{\mathcal{A}}}
\newcommand{\SSym}{\tilde{\mathcal{S}}}
\newcommand{\Sv}{{\mathcal{A}_7}}
\newcommand{\Sev}{\tilde\Sv}
\newcommand{\LSev}{\tilde{\Lambda}\!_\Sv}
\newcommand{\LSv}{\Lambda\!_\Sv}
\newcommand{\TSv}{T^{\Sv}_{\mathcal W}}
\newcommand{\HSv}{\mathfrak H_7}
\newcommand{\Id}{\op{\mathbbm{1}}}% {\openone}
\newcommand{\op}{}

\title{Quantum entanglement and contextuality with
 complexifications of $E_8$ root system}
\date{}

\author{Alexander Yu.\ Vlasov}%\email{qubeat@mail.ru}%   

\begin{document}

\sloppy

\maketitle

\begin{abstract}
The Witting configuration with 40 complex rays was suggested as a possible reformulation
of Penrose model with two spin-3/2 systems based on geometry of dodecahedron 
and used for analysis of nonlocality and contextuality in quantum mechanics.
Yet another configuration with 120 quantum states is considered in presented work. 
Despite of different number of states both configurations can be derived from 
complexification of 240 minimal vectors of 8D real lattice corresponding to root
system of Lie algebra $E_8$. An analysis of properties of suggested configuration
of quantum states is provided using many analogies with properties of
Witting configuration.
\end{abstract} 

\section{Introduction}
\label{Sec:intro}

The {\em Witting configuration} with 40 quantum states in 4D Hilbert 
space is useful for analysis of contextuality and quantum entanglement \cite{WA,dod}. 
The configuration was discussed earlier \cite{WA} as a convenient reformulation of 
a model based on geometry of dodecahedron with two entangled 
spin-3/2 system introduced by Penrose \cite{ZP,shadows}.
There are 40 bases (tetrads with orthonormal states) important for analysis of quantum 
contextuality in such configuration \cite{WA,dod,ZP,shadows}.

The configuration can be derived from complex Witting polytope introduced by Coxeter \cite{CoxReg}.
The Witting polytope has 240 vertexes corresponding to only 40 different quantum states or rays 
in complex space due to six common phase multipliers 
$\me^{\mi\pi k/3}$, $k=0,\ldots,5$.

Symmetry group of Witting configuration is a discrete subgroup of unitary group $SU(4)$ 
with 51840 transformations \cite{dod,CoxReg}.
It may be useful to consider other discrete subgroups of $SU(4)$ as a possible tool for
construction of similar configurations. Classification of such subgroups may be found in 
\cite{DSU4}. The symmetries of Witting configuration correspond to the subgroup VI from the 
list and the subgroup investigated in presented work has index IV.

Such a group with 5040 transformations is a 4D unitary representation of 
a double cover of alternating group $A_7$ \cite{DSU4}, {\em i.e.}, group of 2520 even 
permutations of seven elements.
The unitary equivalent representation of the group with different choice of generators 
is denoted $\Sev$ in \Sec{2A7}.

For such representation $\Sev$ is the symmetry group of a configuration $\LSev$
with 240 complex vectors introduced in \Sec{States}. The configuration has some 
analogy with Witting polytope,
because in both cases natural maps into configuration of 240 8D real vectors 
correspond to the root system of Lie algebra $E_8$ after some orthogonal transformation.
Despite of the equivalence in 8D real space the complex configurations provide quite 
different models of system with quantum states in 4D Hilbert space.
The Witting configuration includes 40 rays or quantum states with 40 different
bases (orthogonal tetrads) and $\LSev$ corresponds to 120 states and 210 bases.

Construction and table with 210 bases of $\LSev$ are discussed in \Sec{Bas}
together with problem of contextuality. Entanglement and measurement of
two configurations have some similarity with properties of Witting configuration 
\cite{dod} and analysed in \Sec{Ent}.
The final section summarizes the results of this work and mentions some unaddressed topics.

\section{Group $\Sev$}
\label{Sec:2A7}

Let us consider two matrices
\begin{equation}\label{gen2a4}
\op P_1 = \begin{pmatrix}
1 & 0 & 0 & 0\\
0 & 0 & 0 & 1\\
0 & 1 & 0 & 0\\
0 & 0 & 1 & 0
\end{pmatrix},\quad
\op P_2 = \begin{pmatrix}
\ 0 &\ 0 &\ 1 &\ 0\\
\ 0 &\ 0 &\ 0 &\ 1\\
-1  &\ 0 &\ 0 &\ 0\\
\ 0 & -1 &\ 0 &\ 0
\end{pmatrix}.
\end{equation}
The matrices $\op P_1$ and $\op P_2$ generate group of 24 transformations isomorphic
with double cover $\SAlt_4$ of group $A_4$ of even permutation of {\em four elements}
in quite natural way as permutations of four basic states.

An isomorphism of the 4D unitary representation of group $\Sev$ of 5040 transformations 
with double cover of group $A_7$ of even permutation of {\em seven elements} is less trivial, 
but it can be generated by $\op P_1$ and $\op P_2$ together with matrix of seventh order
\begin{equation}\label{gen3a7}
\op S = \begin{pmatrix}
1 & 0 & 0 & 0\\
0 & 2\iota & \iota' & \iota\\
0 & \iota' & \iota & -\iota''\\
0 & \iota & -\iota'' & -\cc\iota'
\end{pmatrix},
\end{equation}
where
\begin{equation}
\label{iots}
  \iota = \frac{\mi}{\sqrt{7}},\quad
  \iota' = \frac{1+\iota}{2}, \quad
  \iota'' = \frac{1+3\iota}{2}. 
\end{equation}
Let us note $|\iota|^2=\dfrac{1}{7}$, $|\iota'|^2=\dfrac{2}{7}$ 
and $|\iota''|^2=\dfrac{4}{7}$.

\smallskip

Formally the symmetries of quantum states as rays in Hilbert space
could be described more naturally by group of 2520 {\em projective transformations} 
$\Sv = \Sev/\{+\Id,-\Id\}$. The group
$\Sv$ is isomorphic with group $A_7$ of even permutations of set $\{1,\ldots,7\}$
already mentioned earlier.
With system of computer algebra GAP \cite{GAP}
the 2-1 homomorphism $\HSv : \Sev \mapsto A_7$ can be defined on generators as
\begin{equation}
\label{HSv}
\begin{array}{rccl}
\HSv : &\pm \op P_1 &\to& (1,2,4)(3,6,5), \\
\HSv : &\pm \op P_2 &\to& (1,6)(3,4), \\ 
\HSv : &\pm \op S   &\to&  (1,2,3,4,5,6,7).
\end{array}
\end{equation}
Due to homomorphism $\HSv$ any projective transformation from
$\Sv$ can be naturally encoded into even permutation of
$\{1,\ldots,7\}$.

\section{States}
\label{Sec:States}

The group $\Sev$ represents symmetries of 4D complex configuration $\LSev$ with 240
vectors. The configuration can be produced from a basic vector $(1,0,0,0)$ by application
different transformations from $\Sev$. Each vector is included together with an opposite 
one and it corresponds to configuration $\LSv$ with 120 different quantum states represented
on Table~\ref{tab120} with notation for coordinates from \Eq{iots} and 
indexing explained below. Complex Witting polytope
also has 240 vertices, but only 40 {\em different} quantum states can be obtained
from such vectors due to six common phase multipliers $\pm\omega^k$, $k=0,1,2$, where
\begin{equation}
\label{omg}
 \omega=\me^{\frac{2\mi\pi}{3}}= \frac{-1+ \sqrt{3}\,\mi}{2}.
\end{equation}

\begin{table}[htbp]
\caption{Coefficients and indexes of 120 states.}
\label{tab120} 
\[
{%\setlength{\arraycolsep}{0.1em}
\begin{array}{|r||c|c|c|c||r|}
\hline
n&\ket{\psi_n}&\ket{\psi_{n+10}}&\ket{\psi_{n+20}}&\ket{\psi_{n+30}}&t\\
\hline
1& (1,0,0,0)&
(0,1,0,0)&
(0,0,1,0)&
(0,0,0,1)&1\\
2& (2\iota,\iota,\iota,\iota)&
(\iota,-2\iota,-\iota,\iota)&
(\iota,\iota,-2\iota,-\iota)&
(\iota,-\iota,\iota,-2\iota)&2\\
3& (\iota,0,\cc\iota',\cc\iota'')&
(\iota',\iota,0,\iota'')&
(\cc\iota',\cc\iota'',-\iota,0)&
(0,-\iota'',\iota',\iota)&4\\
4& (\iota,0,-\iota',-\iota'')&
(\cc\iota',-\iota,0,\cc\iota'')&
(\iota',\iota'',\iota,0)&
(0,\cc\iota'',-\cc\iota',\iota)&4\\
5& (0,\cc\iota',-\iota,\iota'')&
(\cc\iota',0,-\iota'',-\iota)&
(\iota,-\iota'',0,\cc\iota')&
(\iota'',\iota,\cc\iota',0)&4\\
6& (\iota,-\cc\iota',-\cc\iota',\iota')&
(\iota',-\iota,\cc\iota',-\cc\iota')&
(\cc\iota',-\iota',\iota,-\cc\iota')&
(\cc\iota',-\cc\iota',-\iota',\iota)&3\\
7& (\iota,-\iota',\cc\iota'',0)&
(0,\iota,\cc\iota'',\iota')&
(\cc\iota'',0,-\iota,\iota')&
(\cc\iota'',\iota',0,-\iota)&4\\
8& (0,\iota',\iota,\cc\iota'')&
(\iota',0,-\cc\iota'',\iota)&
(\iota,\cc\iota'',0,-\iota')&
(\cc\iota'',-\iota,\iota',0)&4\\
9& (\iota,\cc\iota',-\iota'',0)&
(0,\iota,-\iota'',-\cc\iota')&
(\iota'',0,\iota,\cc\iota')&
(\iota'',\cc\iota',0,\iota)&4\\
10& (\iota,\iota',\iota',-\cc\iota')&
(\cc\iota',\iota,\iota',-\iota')&
(-\iota',\cc\iota',\iota,\iota')&
(-\iota',\iota',\cc\iota',\iota)&3\\
\hline\hline
41& (\cc\iota',2\iota,-\iota,0)&
(2\iota,-\cc\iota',0,-\iota)&
(\iota,0,\cc\iota',2\iota)&
(0,\iota,2\iota,-\cc\iota')&4\\
42& (\iota,\iota',-\cc\iota',\iota')&
(-\iota',\iota,\iota',\cc\iota')&
(\cc\iota',-\iota',\iota,\iota')&
(\iota',\cc\iota',\iota',-\iota)&3\\
43& (\iota,\cc\iota',\cc\iota'',0)&
(0,\iota,-\iota'',\iota')&
(\cc\iota'',0,-\iota,-\cc\iota')&
(\iota'',-\iota',0,\iota)&4\\
44& (0,\iota',\iota,-\iota'')&
(\cc\iota',0,\cc\iota'',-\iota)&
(-\iota,\iota'',0,\iota')&
(-\cc\iota'',\iota,\cc\iota',0)&4\\
45& (\iota,\iota,\iota'',-\iota)&
(\iota,-\iota,\iota,\iota'')&
(-\iota'',\iota,\iota,\iota)&
(\iota,\iota'',-\iota,\iota)&2\\
46& (\iota,0,-\iota',\cc\iota'')&
(\iota',\iota,0,-\cc\iota'')&
(\iota',-\cc\iota'',\iota,0)&
(0,\cc\iota'',\iota',\iota)&4\\
47& (2\iota,0,-\iota,-\cc\iota')&
(\cc\iota',0,2\iota,-\iota)&
(\iota,\cc\iota',2\iota,0)&
(2\iota,-\iota,-\cc\iota',0)&4\\
48& (-\iota,\cc\iota',\cc\iota',\cc\iota')&
(\cc\iota',\iota,-\cc\iota',\cc\iota')&
(\cc\iota',\cc\iota',\iota,-\cc\iota')&
(\cc\iota',-\cc\iota',\cc\iota',\iota)&3\\
49& (\iota,-\cc\iota',\iota',\iota')&
(\iota',-\iota,\cc\iota',\iota')&
(\iota',\iota',-\iota,\cc\iota')&
(\cc\iota',\iota',-\iota',\iota)&3\\
50& (0,\iota',\iota,2\iota)&
(0,2\iota,\iota',\iota)&
(\iota,2\iota,0,-\iota')&
(\iota',\iota,0,-2\iota)&4\\
\hline\hline
81& (\iota',-2\iota,\iota,0)&
(2\iota,\iota',0,-\iota)&
(\iota,0,-\iota',2\iota)&
(0,\iota,2\iota,\iota')&4\\
82& (\iota,-\cc\iota',\iota',-\cc\iota')&
(\cc\iota',\iota,-\cc\iota',-\iota')&
(\iota',-\cc\iota',-\iota,\cc\iota')&
(\cc\iota',\iota',\cc\iota',\iota)&3\\
83& (0,\cc\iota',-\iota,-\cc\iota'')&
(\iota',0,\iota'',\iota)&
(\iota,\cc\iota'',0,\cc\iota')&
(\iota'',\iota,-\iota',0)&4\\
84& (-\iota,\iota',\iota'',0)&
(0,\iota,\cc\iota'',-\cc\iota')&
(\iota'',0,\iota,-\iota')&
(\cc\iota'',-\cc\iota',0,-\iota)&4\\
85& (\iota,\iota',\iota',\iota')&
(\iota',-\iota,-\iota',\iota')&
(\iota',\iota',-\iota,-\iota')&
(\iota',-\iota',\iota',-\iota)&3\\
86& (0,-\cc\iota',\iota,2\iota)&
(0,2\iota,-\cc\iota',\iota)&
(\iota,2\iota,0,\cc\iota')&
(\cc\iota',-\iota,0,2\iota)&4\\
87& (\iota,\iota',-\cc\iota',-\cc\iota')&
(\cc\iota',\iota,\iota',\cc\iota')&
(\cc\iota',\cc\iota',\iota,\iota')&
(\iota',\cc\iota',-\cc\iota',-\iota)&3\\
88& (\iota,\iota,-\cc\iota'',-\iota)&
(\iota,-\iota,\iota,-\cc\iota'')&
(\cc\iota'',\iota,\iota,\iota)&
(\iota,-\cc\iota'',-\iota,\iota)&2\\
89& (2\iota,0,-\iota,\iota')&
(\iota',0,-2\iota,\iota)&
(\iota,-\iota',2\iota,0)&
(2\iota,-\iota,\iota',0)&4\\
90& (\iota,0,\cc\iota',-\iota'')&
(-\cc\iota',\iota,0,\iota'')&
(-\cc\iota',\iota'',\iota,0)&
(0,\iota'',\cc\iota',-\iota)&4\\
\hline
\end{array}
}
\]
\end{table}

Despite of such difference both configurations considered as 240 8D real
vectors up to some orthogonal transformation are equivalent with $E_8$ root 
system or minimal vectors of $E_8$ lattice
denoted further simply as $E_8$.
For Witting polytope relation with the 8D real uniform polytope $4_{21}$ (equivalent 
with $E_8$) was mentioned by Coxeter \cite{CoxReg,Cox88}.
%Maps between realifications of both complex configurations with 240 vectors
%and $E_8$ can be constructed using GAP \cite{GAP}. 
The construction of a map between realifications of both complex configurations is 
provided below.

For certainty let us fix 40 vectors producing whole Witting configuration by multiplication
on six `phases' $\pm\omega^k$ \Eq{omg},  $k = 0,1,2$ 
\begin{subequations}\label{psi40}
\begin{equation}\label{psi40bas}
 (1,0,0,0),\qquad (0,1,0,0),\qquad
 (0,0,1,0),\qquad (0,0,0,1),
\end{equation}
\begin{equation}
\begin{split}\label{psi40rest}
 \frac{\mi}{\sqrt{3}}(0,1,-\omega^\mu,\omega^\nu),\quad
 &\frac{\mi}{\sqrt{3}}(1,0,-\omega^\mu,-\omega^\nu), \\
 &\frac{\mi}{\sqrt{3}}(1,-\omega^\mu,0,\omega^\nu),\quad
 \frac{\mi}{\sqrt{3}}(1,\omega^\mu,\omega^\nu,0)
\end{split} 
\end{equation}
\end{subequations}
with $\mu, \nu = 0,1,2$.

Let us note, that four vectors \Eq{psi40rest} with $\mu = \nu = 0$ have
pure imaginary coordinates 
\begin{equation}
\begin{split}\label{psi4Im}
 \frac{\mi}{\sqrt{3}}(0,1,-1,1),\quad
 &\frac{\mi}{\sqrt{3}}(1,0,-1,-1), \\
 &\frac{\mi}{\sqrt{3}}(1,-1,0,1),\quad
 \frac{\mi}{\sqrt{3}}(1,1,1,0)
\end{split} 
\end{equation}
and together with \Eq{psi40bas} can be used
as convenient basis for realification of Witting polytope. 
$\LSev$ also includes four states \Eq{psi40bas} and four vectors with pure 
imaginary components
\begin{equation}\label{LSevIm}
\begin{split}
 \frac{\mi}{\sqrt{7}}(2,1,1,1),\quad
 &\frac{\mi}{\sqrt{7}}(1,-2,-1,1), \\
 &\frac{\mi}{\sqrt{7}}(1,1,-2,-1),\quad
 \frac{\mi}{\sqrt{7}}(1,-1,1,-2)
\end{split} 
\end{equation} 

After some technical manipulations with GAP \cite{GAP}
a simple transformation of Witting polytope into $\LSev$ can be found.
It saves all real 
parts of coordinates and multiply four-vectors with imaginary 
parts on the orthogonal matrix 
\begin{equation}
\TSv =  \frac{1}{\sqrt{21}}\begin{pmatrix}
\ 1&\ 4& -2&\ 0\\
 -4&\ 1&\ 0&\ 2\\
\ 2&\ 0&\ 1&\ 4\\
\ 0& -2& -4&\ 1\\
\end{pmatrix}.
\end{equation}
The transformation also converts \Eq{psi4Im} into \Eq{LSevIm}.

Let us use indexation of 40 states \Eq{psi40} in Witting configuration introduced 
(up to normalization) in Ref.~\cite{dod} to produce 120 vectors by appending two extra blocks
multiplied on $\omega$ \Eq{omg} and $\omega^2$ respectively. All 120 states 
$\LSv$ can be obtained now by transformation of imaginary parts of all vectors using $\TSv$. 

The result of such operation is presented in Table~\ref{tab120}. 
The choice between two possible directions of a vector is often due to
some typographical reasons.
All states are normalized on unit and two copies of Table~\ref{tab120} with opposite 
signs produces complete set $\LSev$ with 240 complex vectors.

Index $t$ in last column of Table~\ref{tab120} denotes {\em type} of four states in a row similar with
used for description of lattice for group $\Sev$ denoted as $\mathtt{2A_7}$ in Ref.~\cite[p.~10]{Atl}
and up to 24 `signed' permutations from $\SAlt_4$ it corresponds to  
\begin{enumerate}
 \item 4 basic states (8 vectors) from $(1,0,0,0)$
 \item 12 states (24 vectors) from $(a b,1,1,1)\cdot\mi/\sqrt{7}$ 
 \item 32 states (64 vectors) from $(1,-a,-b,-c)\cdot\mi/\sqrt{7}$
 \item 72 states (144 vectors) from $(0,1,-a b,c)\cdot\mi/\sqrt{7}$
\end{enumerate}
Here $a,b,c = \dfrac{-1\pm\mi\sqrt{7}}2$, {\em i.e.}, the expressions include all possible 
combinations with two complex roots of quadratic equation $z^2+z+2=0$ \cite{E8}.
Let us note, the $\omega$ and $\cc\omega$ used for construction of Witting polytope
are roots of quadratic equation $z^2+z+1=0$.

\section{Bases and contextuality}
\label{Sec:Bas}

Number of bases (orthonormal tetrads) for the 120 states is 210. Consequent numbering of
such bases in lexicographic order with respect to indexes of states is shown in 
Table~\ref{tab210}. Each state belongs to seven different bases collecting
all 21 states orthogonal to it. A diagram with links corresponding to orthogonal
states is known as Kochen-Specker graph \cite{WA,ZP}. All 210 bases can be found
using search for maximal 4-cliques in the graph.

\begin{table}[htbp]
\caption{210 bases with states from Table \ref{tab120}.}
\label{tab210}
{\setlength{\arraycolsep}{0.2em} 
\renewcommand{\arraystretch}{0.95}
\[
\begin{array}{|r|l||r|l||r|l|}
\hline
 n & B_n & n & B_n & n & B_n \\ \hline
1 & (1, 5, 33, 111) & 2 & (1, 8, 34, 71) & 3 & (\ul{1, 11, 21, 31}) \\
4 & (1, 17, 83, 96) & 5 & (1, 19, 44, 60) & 6 & (1, 50, 53, 120) \\
7 & (1, 76, 86, 94) & 8 & (2, 7, 26, 83) & 9 & (2, 9, 30, 44) \\
10 & (\ul{2, 12, 22, 32}) & 11 & (2, 25, 52, 53) & 12 & (2, 28, 92, 94) \\
13 & (2, 33, 40, 90) & 14 & (2, 34, 36, 46) & 15 & (3, 10, 77, 118) \\
16 & (3, 11, 27, 57) & 17 & (3, 12, 38, 82) & 18 & (3, 24, 78, 95) \\
19 & (3, 39, 53, 66) & 20 & (3, 58, 81, 112) & 21 & (\ul{3, 73, 75, 88}) \\
22 & (4, 6, 75, 119) & 23 & (4, 11, 29, 99) & 24 & (4, 12, 35, 42) \\
25 & (4, 23, 58, 115) & 26 & (4, 37, 94, 110) & 27 & (4, 41, 72, 95) \\
28 & (\ul{4, 45, 114, 118}) & 29 & (5, 7, 91, 97) & 30 & (5, 18, 106, 119) \\
31 & (5, 22, 72, 73) & 32 & (5, 28, 54, 113) & 33 & (\ul{5, 36, 38, 49}) \\
34 & (5, 37, 52, 67) & 35 & (6, 15, 56, 81) & 36 & (6, 17, 30, 39) \\
37 & (6, 22, 27, 103) & 38 & (\ul{6, 26, 59, 79}) & 39 & (6, 50, 74, 78) \\
40 & (6, 65, 82, 87) & 41 & (\ul{7, 14, 104, 120}) & 42 & (7, 20, 39, 86) \\
43 & (7, 23, 31, 77) & 44 & (7, 29, 60, 116) & 45 & (7, 62, 100, 112) \\
46 & (8, 9, 51, 59) & 47 & (8, 15, 70, 77) & 48 & (8, 22, 112, 114) \\
49 & (8, 25, 74, 93) & 50 & (\ul{8, 35, 40, 87}) & 51 & (8, 39, 92, 109) \\
52 & (\ul{9, 13, 63, 76}) & 53 & (9, 16, 37, 50) & 54 & (9, 24, 31, 119) \\
55 & (9, 27, 80, 96) & 56 & (9, 56, 72, 102) & 57 & (10, 18, 41, 100) \\
58 & (10, 19, 26, 37) & 59 & (10, 22, 29, 64) & 60 & (\ul{10, 30, 97, 117}) \\
61 & (10, 42, 49, 108) & 62 & (10, 86, 113, 115) & 63 & (11, 15, 101, 104) \\
64 & (11, 18, 61, 63) & 65 & (11, 46, 47, 54) & 66 & (11, 89, 90, 93) \\
67 & (12, 17, 63, 107) & 68 & (12, 19, 69, 104) & 69 & (12, 54, 56, 117) \\
70 & (12, 79, 93, 100) & 71 & (13, 20, 22, 110) & 72 & (13, 21, 25, 91) \\
73 & (13, 34, 65, 88) & 74 & (13, 40, 67, 105) & 75 & (13, 43, 68, 115) \\
76 & (13, 45, 81, 92) & 77 & (14, 16, 22, 66) & 78 & (14, 21, 28, 51) \\
79 & (14, 33, 45, 108) & 80 & (14, 36, 68, 109) & 81 & (14, 41, 52, 88) \\
82 & (14, 78, 84, 105) & 83 & (\ul{15, 20, 28, 107}) & 84 & (15, 24, 32, 62) \\
85 & (15, 38, 64, 83) & 86 & (15, 42, 66, 116) & 87 & (\ul{16, 18, 25, 69}) \\
88 & (16, 34, 48, 89) & 89 & (16, 40, 49, 107) & 90 & (16, 55, 60, 113) \\
91 & (16, 75, 92, 97) & 92 & (17, 24, 90, 114) & 93 & (17, 25, 47, 72) \\
94 & (17, 29, 36, 70) & 95 & (\ul{17, 51, 101, 110}) & 96 & (18, 23, 32, 102) \\
97 & (18, 35, 44, 103) & 98 & (18, 80, 82, 110) & 99 & (19, 23, 46, 73) \\
100 & (19, 27, 40, 106) & 101 & (19, 28, 89, 112) & 102 & (\ul{19, 61, 66, 91}) \\
103 & (20, 33, 47, 85) & 104 & (20, 36, 69, 87) & 105 & (20, 52, 59, 118) \\
\hline 
\end{array}
\]
}
\end{table} 

\begin{table}[htbp]
\begin{center}
Table \ref{tab210} (continued): 210 bases with states from Table \ref{tab120}.
{\setlength{\arraycolsep}{0.2em} 
\renewcommand{\arraystretch}{0.95}
\[
\begin{array}{|r|l||r|l||r|l|}
\hline
 n & B_n & n & B_n & n & B_n \\ \hline
106 & (20, 74, 96, 98) & 107 & (21, 37, 103, 116) & 108 & (21, 39, 64, 80) \\
109 & (21, 56, 106, 114) & 110 & (21, 70, 73, 100) & 111 & (23, 30, 57, 98) \\
112 & (\ul{23, 53, 55, 108}) & 113 & (23, 78, 92, 101) & 114 & (24, 26, 55, 99) \\
115 & (24, 52, 61, 115) & 116 & (\ul{24, 65, 94, 98}) & 117 & (25, 27, 111, 117) \\
118 & (25, 38, 86, 99) & 119 & (26, 35, 76, 101) & 120 & (26, 45, 102, 107) \\
121 & (26, 54, 58, 70) & 122 & (\ul{27, 34, 84, 100}) & 123 & (27, 42, 92, 120) \\
124 & (28, 29, 71, 79) & 125 & (28, 35, 50, 57) & 126 & (\ul{29, 33, 43, 56}) \\
127 & (29, 52, 76, 82) & 128 & (30, 38, 61, 120) & 129 & (30, 62, 69, 88) \\
130 & (30, 93, 95, 106) & 131 & (31, 35, 81, 84) & 132 & (31, 38, 41, 43) \\
133 & (31, 66, 67, 74) & 134 & (31, 109, 110, 113) & 135 & (32, 37, 43, 87) \\
136 & (32, 39, 49, 84) & 137 & (32, 59, 113, 120) & 138 & (32, 74, 76, 97) \\
139 & (33, 48, 63, 95) & 140 & (33, 65, 101, 112) & 141 & (34, 58, 85, 104) \\
142 & (34, 61, 72, 108) & 143 & (35, 46, 62, 96) & 144 & (36, 55, 112, 117) \\
145 & (36, 75, 80, 93) & 146 & (\ul{37, 71, 81, 90}) & 147 & (38, 60, 90, 102) \\
148 & (\ul{39, 41, 46, 111}) & 149 & (40, 54, 116, 118) & 150 & (40, 72, 79, 98) \\
151 & (41, 57, 85, 116) & 152 & (41, 71, 91, 101) & 153 & (42, 43, 51, 98) \\
154 & (\ul{42, 52, 102, 112}) & 155 & (42, 71, 73, 105) & 156 & (43, 55, 104, 118) \\
157 & (43, 78, 80, 107) & 158 & (\ul{44, 48, 74, 115}) & 159 & (44, 54, 75, 108) \\
160 & (44, 65, 107, 109) & 161 & (44, 77, 79, 85) & 162 & (45, 50, 73, 117) \\
163 & (45, 64, 87, 89) & 164 & (45, 98, 103, 113) & 165 & (46, 79, 97, 110) \\
166 & (46, 107, 119, 120) & 167 & (47, 49, 88, 103) & 168 & (\ul{47, 60, 67, 80}) \\
169 & (47, 71, 106, 115) & 170 & (47, 76, 100, 109) & 171 & (48, 53, 105, 114) \\
172 & (48, 80, 81, 99) & 173 & (48, 83, 117, 119) & 174 & (48, 91, 94, 102) \\
175 & (49, 56, 57, 110) & 176 & (49, 95, 96, 104) & 177 & (50, 58, 91, 109) \\
178 & (\ul{50, 77, 99, 106}) & 179 & (51, 53, 62, 85) & 180 & (51, 61, 81, 111) \\
181 & (51, 67, 86, 95) & 182 & (53, 65, 70, 97) & 183 & (\ul{54, 64, 68, 95}) \\
184 & (55, 64, 74, 88) & 185 & (55, 82, 84, 91) & 186 & (56, 67, 89, 120) \\
187 & (57, 59, 64, 105) & 188 & (\ul{57, 70, 86, 119}) & 189 & (58, 60, 63, 87) \\
190 & (\ul{58, 93, 103, 105}) & 191 & (59, 66, 90, 117) & 192 & (59, 94, 106, 108) \\
193 & (60, 68, 101, 119) & 194 & (61, 77, 96, 105) & 195 & (62, 63, 71, 118) \\
196 & (\ul{62, 72, 82, 92}) & 197 & (63, 75, 84, 98) & 198 & (65, 83, 93, 118) \\
199 & (66, 87, 99, 100) & 200 & (67, 69, 83, 108) & 201 & (68, 73, 85, 94) \\
202 & (68, 82, 111, 114) & 203 & (68, 97, 99, 103) & 204 & (69, 76, 77, 90) \\
205 & (69, 84, 115, 116) & 206 & (70, 78, 89, 111) & 207 & (75, 102, 104, 111) \\
208 & (\ul{78, 83, 85, 113}) & 209 & (79, 86, 88, 114) & 210 & (\ul{89, 96, 109, 116}) \\
\hline
\end{array}
\]
}
\end{center}
\end{table} 

Yet another method used earlier in Ref.~\cite{dod} for Witting configuration is due to 
relation between bases and $4 \times 4$ unitary matrices representing elements of symmetry group
$\Sev$. Each bases corresponds to $5040/210=24$ different elements of $\Sev$ and such
ambiguity corresponds to 24 permutations and changes directions of vectors in the bases
described by group $\SAlt_4$ .
Unlike Witting configuration there is no subgroup of symmetry group useful for
creation of full list with 210 bases in such a way.

Let us consider contextuality problem for given configuration of quantum states 
$\LSv$ with a method used earlier for Witting configuration \cite{dod}. A noncontextual 
classical model reproducing 
principles of quantum mechanics would require a map of 120 
vectors into $\{0,1\}$ with property: {\em one and only one vector for any basis
is mapped into 1}. 

Let us consider some partition of 120 states on 30 bases, {\em e.g.},
30 bases underlined in Table~\ref{tab210}. All 30 states mapped into 1 in such
partitions should be non-orthogonal, because any two orthogonal states
in the $\LSv$ could be extended into orthogonal basis and it would contain 
more than one state mapped into one.

Thus, the noncontextual model would require existence of 30 mutually nonorthogonal states. 
A brute-force search used in Ref.~\cite{dod} for Witting configuration 
supposes to find all maximal non-orthogonal cliques. Such a method is not effective, but still can be
used for $\LSv$. The list with sizes of all such cliques obtained with GAP software \cite{GAP} 
is presented below.
\begin{equation}
\label{nncl}
{\setlength{\arraycolsep}{0.2em}
\begin{array}{r}
\begin{array}{|r||c|c|c|c|c|c|c|}
\hline
\text{Size}&   10&    11&     12&       13&      14&      15&    16\\ \hline
\text{Num.}& 3528& 15120& 277130& 1117200& 3802620& 7018440& 12077100\\
\hline 
\end{array}\ \\ \hookleftarrow\\
\setlength{\arraycolsep}{0.3em}
\hookrightarrow\begin{array}{|r||c|c|c|c|c|c|c|c|}
\hline
\text{Size} &      17&      18&     19&     20&   21&    22& 23& 24\\ \hline
\text{Num.}& 3619560& 2737980& 446040& 342636& 1120& 16860& 0 & 420\\
\hline
\end{array}
\end{array}
}
\end{equation}
The nonorthogonal cliques with size 30 do not exist and maximal size is 24.

\section{Entanglement of two configurations}
\label{Sec:Ent}

Let us consider an entangled state
\begin{equation}\label{entOm}
\ket{\Omega}=\frac{1}{2}\bigl(\ket{0}\ket{3}-\ket{1}\ket{2}+\ket{2}\ket{1}-\ket{3}\ket{0}\bigr).
\end{equation} 
In more general case an entangled state can be described by some matrix $\op J$ as
\begin{equation}\label{EntJ}
 \ket{\Omega_J} = \frac{1}{\sqrt{\mathrm{Tr}(\op J\!\op J^*)}} \sum_{jk} J_{jk}\ket{j}\ket{k}.  
\end{equation} 
For state $\ket{\Omega}$ in \Eq{entOm} 
\begin{equation}\label{J0}
 \op J = \begin{pmatrix}
  0 & 0 & 0 & 1\\
  0 & 0 & -1& 0\\
  0 & 1 & 0 & 0\\
  -1& 0 & 0 & 0
 \end{pmatrix}.
\end{equation} 

For consideration of different scenarios of measurement with such kind of configurations 
it is useful to consider transformations of both systems respecting state \Eq{EntJ}.
A natural example \cite{dod} is a measurement bases for both systems obtained by the 
same unitary transformation $\op A$ with property 
\begin{equation}\label{AJpmJA}
  (\op A \otimes \op A) \ket{\Omega_J} = \pm \ket{\Omega_J} \quad\Longrightarrow\quad
  \cc{\op{A}} \op J = \pm \op J \op A, 
\end{equation} 
where $\cc{\op{A}}$ is complex conjugation of all coefficients in $\op A$. 
It is enough to check property \Eq{AJpmJA} for generators of some subgroup, because
products also would respect the same relation.  
For matrix \Eq{J0} there are subgroups of $\Sev$ with transformations satisfying \Eq{AJpmJA}
and respecting entangled state \Eq{entOm}. 
The first group is $\SAlt_5$ with 120 transformation generated by matrix $\op J$ together with
\begin{equation}\label{gen2a5}
\op H = \begin{pmatrix}
0 & \iota' & \iota & -\iota''\\
\iota' & -\iota' & \iota' & -\iota\\
\iota & -\cc\iota' & -\cc\iota' & -\cc\iota'\\
\cc\iota'' & -\iota & -\cc\iota' & 0
\end{pmatrix},
\end{equation}
Let us note analogues of maps \Eq{HSv} for generators of the subgroup
\begin{equation}
\label{HSvJH}
\HSv : \pm \op J \to (1,6)(2,5), \quad
\HSv : \pm \op H \to (1,6,5,2,7).
\end{equation}
The second subgroup is $\SSym_5$ with 240 transformations generated by
$\op J$, $\op H$ and $\op P_2$. The subgroups $\SAlt_5$ and $\SSym_5$ are double covers 
of $A_5$ and $S_5$ representing even and all permutations of five elements 
respectively.

Symmetries of Witting configuration also include similar bigger subgroups with
720 and 1440 transformations useful for analysys of entanglement \cite{dod}.
However for Witting configuration both such subgroups act transitively on all
40 vectors, but for $\LSv$ only orbit of $\SSym_5$ includes all 120 vectors
and $\SAlt_5$ has two orbits with 60 vectors.
The first orbit includes basic vectors $\ket{0}$ and $\ket{3}$ and the second
one $\ket{1}$ and $\ket{2}$.
Despite only group  $\SSym_5$ includes orbit with whole system of 120 states,
both groups can be used for construction from the four basic states 
$\ket{0},\ldots,\ket{3}$ all 30 bases underlined in Table~\ref{tab210}.
 
Let us consider two scheme of measurements of entangled systems similar with
discussed in Ref.~\cite{dod}. In a simpler one measurement bases for both systems 
are chosen by application of some transformation from $\SAlt_5$ or $\SSym_5$. 
However in such a case only 30 bases between 210
can be used. It is similar with analogue property of Witting configuration
with only 10 bases between 40 are available for equivalent measurement of
both systems.

Let us also recollect construction of `$J$-opposite state' defined by anti-unitary
transformation up to unsignificant phase \cite{dod} 
\begin{equation}\label{Jopp}
  \ket{\psi_J} \simeq \op J \ket{\cc{\psi}}.
\end{equation}
The result of measurement for entangled state \Eq{entOm} is always pair of such 
`$J$-opposite states'. The indexes for such pairs for configuration $\LSv$ are 
collected in Table~\ref{TabOppJ}.  

\begin{table}[htb]
\caption{Pairs of indexes for $J$-opposite states.}
\label{TabOppJ}
\begin{center}
{\setlength{\arraycolsep}{0.15em}
\newcommand{\psep}{\leftrightarrow}
\(
\begin{array}{|rcl|rcl|rcl|rcl|rcl|rcl|}
\hline
 1 &\psep& 31 & 2 &\psep& 32 & 3 &\psep& 73 & 4 &\psep& 114 & 5 &\psep& 38 & 6 &\psep& 79 \\ 
 7 &\psep& 120 & 8 &\psep& 35 & 9 &\psep& 76 & 10 &\psep& 117 & 11 &\psep& 21 & 12 &\psep& 22 \\ 
 13 &\psep& 63 & 14 &\psep& 104 & 15 &\psep& 28 & 16 &\psep& 69 & 17 &\psep& 110 & 18 &\psep& 25 \\ 
 19 &\psep& 66 & 20 &\psep& 107 & 21 &\psep& 11 & 22 &\psep& 12 & 23 &\psep& 53 & 24 &\psep& 94 \\ 
 25 &\psep& 18 & 26 &\psep& 59 & 27 &\psep& 100 & 28 &\psep& 15 & 29 &\psep& 56 & 30 &\psep& 97 \\ 
 31 &\psep& 1 & 32 &\psep& 2 & 33 &\psep& 43 & 34 &\psep& 84 & 35 &\psep& 8 & 36 &\psep& 49 \\ 
 37 &\psep& 90 & 38 &\psep& 5 & 39 &\psep& 46 & 40 &\psep& 87 & 41 &\psep& 111 & 42 &\psep& 112 \\ 
 43 &\psep& 33 & 44 &\psep& 74 & 45 &\psep& 118 & 46 &\psep& 39 & 47 &\psep& 80 & 48 &\psep& 115 \\
 49 &\psep& 36 & 50 &\psep& 77 & 51 &\psep& 101 & 52 &\psep& 102 & 53 &\psep& 23 & 54 &\psep& 64 \\ 
 55 &\psep& 108 & 56 &\psep& 29 & 57 &\psep& 70 & 58 &\psep& 105 & 59 &\psep& 26 & 60 &\psep& 67 \\
 61 &\psep& 91 & 62 &\psep& 92 & 63 &\psep& 13 & 64 &\psep& 54 & 65 &\psep& 98 & 66 &\psep& 19 \\
 67 &\psep& 60 & 68 &\psep& 95 & 69 &\psep& 16 & 70 &\psep& 57 & 71 &\psep& 81 & 72 &\psep& 82 \\ 
 73 &\psep& 3 & 74 &\psep& 44 & 75 &\psep& 88 & 76 &\psep& 9 & 77 &\psep& 50 & 78 &\psep& 85 \\ 
 79 &\psep& 6 & 80 &\psep& 47 & 81 &\psep& 71 & 82 &\psep& 72 & 83 &\psep& 113 & 84 &\psep& 34 \\ 
 85 &\psep& 78 & 86 &\psep& 119 & 87 &\psep& 40 & 88 &\psep& 75 & 89 &\psep& 116 & 90 &\psep& 37 \\ 
 91 &\psep& 61 & 92 &\psep& 62 & 93 &\psep& 103 & 94 &\psep& 24 & 95 &\psep& 68 & 96 &\psep& 109 \\ 
 97 &\psep& 30 & 98 &\psep& 65 & 99 &\psep& 106 & 100 &\psep& 27 & 101 &\psep& 51 & 102 &\psep& 52 \\ 
 103 &\psep& 93 & 104 &\psep& 14 & 105 &\psep& 58 & 106 &\psep& 99 & 107 &\psep& 20 & 108 &\psep& 55 \\ 
 109 &\psep& 96 & 110 &\psep& 17 & 111 &\psep& 41 & 112 &\psep& 42 & 113 &\psep& 83 & 114 &\psep& 4 \\ 
 115 &\psep& 48 & 116 &\psep& 89 & 117 &\psep& 10 & 118 &\psep& 45 & 119 &\psep& 86 & 120 &\psep& 7 \\
 \hline
\end{array} 
\)
}
\end{center}
\end{table}

The bases underlined in Table~\ref{tab210} correspond to examples with two pairs
of $J$-opposite states included in the same basis. The other method of finding
such bases it to consider application of transformations from group $\SSym_5$
to initial basis $\ket{k}$, $k=0,\ldots,3$. In the Table~\ref{tab120} the basic
states have indexes $10k+1$, {\em i.e.}, $(1,11,21,31)$ and initial basis
has index 3 in the Table~\ref{tab210}. 

The measurement with only 30 bases is not appropriate for analysis of contextuality
\cite{dod,ZP} and the second measurement scheme can be used instead. In such a
case different transformations are used for preparation of measurement bases.
For any $\op A \in \Sev$ the transformation of second system $\op B$ 
can be obtained \cite{dod} due equations 
\begin{equation}\label{BJA}
 (\op A \otimes \op B) \ket{\Omega_J} = \ket{\Omega_J} \quad\Longrightarrow\quad
 \op B \op J = \op J \cc{\op A},\quad \op B = \op J \cc{\op A} \op J^{-1}.
\end{equation}
Let us note, that $\op B \in \Sev$ because $\op J \in \Sev$ and 
generators \Eq{gen2a4} and \Eq{gen3a7} as well as any other 
transformation $\op A \in \Sev$ corresponding to a product
of the generators have property $\cc{\op A} \in \Sev$.

Permutations and directions of vectors in measurement bases described by group $\SAlt_4$
are not essential and thus 5040 transformations from $\Sev$ correspond to
210 different measurement bases. All pairs of such bases obtained by 
transformations $\op A$ and $\op B$ from \Eq{BJA} are represented in Table~\ref{TabOppBasJ}.
For any such pair second basis is corresponding to collection of $J$-opposite
states and {\em vice versa\/}. For 30 underlined bases in Table~\ref{tab210}
both indexes in a pair coincide. 

\begin{table}[htbp]
\caption{Pairs of indexes for bases with $J$-opposite states.}
\label{TabOppBasJ}
\begin{center}
{\setlength{\arraycolsep}{0.15em}
\newcommand{\psep}{\leftrightarrow}
\renewcommand{\arraystretch}{0.95}
\(
\begin{array}{|rcl|rcl|rcl|rcl|rcl|rcl|rcl|}
\hline
1 &\psep& 132 & 2 &\psep& 131 & 3 & = & 3 & 4 &\psep& 134 & 5 &\psep& 133 & 6 &\psep& 43\\
7 &\psep& 54 & 8 &\psep& 137 & 9 &\psep& 138 & 10 & = & 10 & 11 &\psep& 96 & 12 &\psep& 84\\
13 &\psep& 135 & 14 &\psep& 136 & 15 &\psep& 162 & 16 &\psep& 110 & 17 &\psep& 31 & 18 &\psep& 201\\
19 &\psep& 99 & 20 &\psep& 155 & 21 & = & 21 & 22 &\psep& 209 & 23 &\psep& 109 & 24 &\psep& 48\\
25 &\psep& 171 & 26 &\psep& 92 & 27 &\psep& 202 & 28 & = & 28 & 29 &\psep& 128 & 30 &\psep& 118\\
31 &\psep& 17 & 32 &\psep& 85 & 33 & = & 33 & 34 &\psep& 147 & 35 &\psep& 124 & 36 &\psep& 165\\
37 &\psep& 70 & 38 & = & 38 & 39 &\psep& 161 & 40 &\psep& 150 & 41 & = & 41 & 42 &\psep& 166\\
43 &\psep& 6 & 44 &\psep& 186 & 45 &\psep& 123 & 46 &\psep& 119 & 47 &\psep& 125 & 48 &\psep& 24\\
49 &\psep& 97 & 50 & = & 50 & 51 &\psep& 143 & 52 & = & 52 & 53 &\psep& 204 & 54 &\psep& 7\\
55 &\psep& 170 & 56 &\psep& 127 & 57 &\psep& 117 & 58 &\psep& 191 & 59 &\psep& 69 & 60 & = & 60\\
61 &\psep& 144 & 62 &\psep& 173 & 63 &\psep& 78 & 64 &\psep& 72 & 65 &\psep& 108 & 66 &\psep& 107\\
67 &\psep& 71 & 68 &\psep& 77 & 69 &\psep& 59 & 70 &\psep& 37 & 71 &\psep& 67 & 72 &\psep& 64\\
73 &\psep& 197 & 74 &\psep& 189 & 75 &\psep& 139 & 76 &\psep& 195 & 77 &\psep& 68 & 78 &\psep& 63\\
79 &\psep& 156 & 80 &\psep& 176 & 81 &\psep& 207 & 82 &\psep& 141 & 83 & = & 83 & 84 &\psep& 12\\
85 &\psep& 32 & 86 &\psep& 101 & 87 & = & 87 & 88 &\psep& 205 & 89 &\psep& 104 & 90 &\psep& 200\\
91 &\psep& 129 & 92 &\psep& 26 & 93 &\psep& 98 & 94 &\psep& 175 & 95 & = & 95 & 96 &\psep& 11\\
97 &\psep& 49 & 98 &\psep& 93 & 99 &\psep& 19 & 100 &\psep& 199 & 101 &\psep& 86 & 102 & = & 102\\
103 &\psep& 157 & 104 &\psep& 89 & 105 &\psep& 120 & 106 &\psep& 160 & 107 &\psep& 66 & 108 &\psep& 65\\
109 &\psep& 23 & 110 &\psep& 16 & 111 &\psep& 182 & 112 & = & 112 & 113 &\psep& 179 & 114 &\psep& 192\\
115 &\psep& 174 & 116 & = & 116 & 117 &\psep& 57 & 118 &\psep& 30 & 119 &\psep& 46 & 120 &\psep& 105\\
121 &\psep& 187 & 122 & = & 122 & 123 &\psep& 45 & 124 &\psep& 35 & 125 &\psep& 47 & 126 & = & 126\\
127 &\psep& 56 & 128 &\psep& 29 & 129 &\psep& 91 & 130 &\psep& 203 & 131 &\psep& 2 & 132 &\psep& 1\\
133 &\psep& 5 & 134 &\psep& 4 & 135 &\psep& 13 & 136 &\psep& 14 & 137 &\psep& 8 & 138 &\psep& 9\\
139 &\psep& 75 & 140 &\psep& 153 & 141 &\psep& 82 & 142 &\psep& 185 & 143 &\psep& 51 & 144 &\psep& 61\\
145 &\psep& 167 & 146 & = & 146 & 147 &\psep& 34 & 148 & = & 148 & 149 &\psep& 163 & 150 &\psep& 40\\
151 &\psep& 206 & 152 &\psep& 180 & 153 &\psep& 140 & 154 & = & 154 & 155 &\psep& 20 & 156 &\psep& 79\\
157 &\psep& 103 & 158 & = & 158 & 159 &\psep& 184 & 160 &\psep& 106 & 161 &\psep& 39 & 162 &\psep& 15\\
163 &\psep& 149 & 164 &\psep& 198 & 165 &\psep& 36 & 166 &\psep& 42 & 167 &\psep& 145 & 168 & = & 168\\
169 &\psep& 172 & 170 &\psep& 55 & 171 &\psep& 25 & 172 &\psep& 169 & 173 &\psep& 62 & 174 &\psep& 115\\
175 &\psep& 94 & 176 &\psep& 80 & 177 &\psep& 194 & 178 & = & 178 & 179 &\psep& 113 & 180 &\psep& 152\\
181 &\psep& 193 & 182 &\psep& 111 & 183 & = & 183 & 184 &\psep& 159 & 185 &\psep& 142 & 186 &\psep& 44\\
187 &\psep& 121 & 188 & = & 188 & 189 &\psep& 74 & 190 & = & 190 & 191 &\psep& 58 & 192 &\psep& 114\\
193 &\psep& 181 & 194 &\psep& 177 & 195 &\psep& 76 & 196 & = & 196 & 197 &\psep& 73 & 198 &\psep& 164\\
199 &\psep& 100 & 200 &\psep& 90 & 201 &\psep& 18 & 202 &\psep& 27 & 203 &\psep& 130 & 204 &\psep& 53\\
205 &\psep& 88 & 206 &\psep& 151 & 207 &\psep& 81 & 208 & = & 208 & 209 &\psep& 22 & 210 & = & 210\\

\hline
\end{array} 
\)
}
\end{center}
\end{table}

Similarly with Witting configuration in Ref.~\cite{dod} alternative choice of entangled states 
and matrices $\op J$ can be obtained by consideration of decomposition of $\Sev$ into cosets
of subgroup $\SSym_5$. There are 5040/240 = 21 different cosets $\op C \SSym_5$ and
alternative entangled states \Eq{EntJ} could be constructed with matrices
\begin{equation}\label{JCT}
 \op J^C = \op C \op J \cc{\op C}^{-1} = \op C \op J \op C^T.
\end{equation}

Except $\op J$ \Eq{J0} only two matrices
between 20 have simplest form with coefficients $0$ and $\pm 1$

\begin{equation}\label{J1}
\op J_1 = \begin{pmatrix}
 \ 0 &\ 0  &\ 1 &\ 0\\
 \ 0 &\ 0  &\ 0 &\ 1\\
\!-1 &\ 0  &\ 0 &\ 0\\
 \ 0 &\!-1 &\ 0 &\ 0
\end{pmatrix}
\end{equation}
and
\begin{equation}\label{J2}
\op J_2 = \begin{pmatrix}
 \ 0 &\ 1 &\ 0 &\ 0\\
\!-1 &\ 0 &\ 0 &\ 0\\
 \ 0 &\ 0 &\ 0 &\!-1\\
 \ 0 &\ 0 &\ 1 &\ 0
\end{pmatrix}.
\end{equation}

\begin{table}[htb]
\caption{Pairs of indexes for $J_1$-opposite states.}
\label{TabOppJ1}
\begin{center}
{\setlength{\arraycolsep}{0.15em}
\newcommand{\psep}{\leftrightarrow}
\(
\begin{array}{|rcl|rcl|rcl|rcl|rcl|rcl|}
\hline
1 &\psep& 11 & 2 &\psep& 12 & 3 &\psep& 53 & 4 &\psep& 94 & 5 &\psep& 18 & 6 &\psep& 59\\
7 &\psep& 100 & 8 &\psep& 15 & 9 &\psep& 56 & 10 &\psep& 97 & 11 &\psep& 1 & 12 &\psep& 2\\
13 &\psep& 43 & 14 &\psep& 84 & 15 &\psep& 8 & 16 &\psep& 49 & 17 &\psep& 90 & 18 &\psep& 5\\
19 &\psep& 46 & 20 &\psep& 87 & 21 &\psep& 31 & 22 &\psep& 32 & 23 &\psep& 73 & 24 &\psep& 114\\
25 &\psep& 38 & 26 &\psep& 79 & 27 &\psep& 120 & 28 &\psep& 35 & 29 &\psep& 76 & 30 &\psep& 117\\
31 &\psep& 21 & 32 &\psep& 22 & 33 &\psep& 63 & 34 &\psep& 104 & 35 &\psep& 28 & 36 &\psep& 69\\
37 &\psep& 110 & 38 &\psep& 25 & 39 &\psep& 66 & 40 &\psep& 107 & 41 &\psep& 91 & 42 &\psep& 92\\
43 &\psep& 13 & 44 &\psep& 54 & 45 &\psep& 98 & 46 &\psep& 19 & 47 &\psep& 60 & 48 &\psep& 95\\
49 &\psep& 16 & 50 &\psep& 57 & 51 &\psep& 81 & 52 &\psep& 82 & 53 &\psep& 3 & 54 &\psep& 44\\
55 &\psep& 88 & 56 &\psep& 9 & 57 &\psep& 50 & 58 &\psep& 85 & 59 &\psep& 6 & 60 &\psep& 47\\
61 &\psep& 111 & 62 &\psep& 112 & 63 &\psep& 33 & 64 &\psep& 74 & 65 &\psep& 118 & 66 &\psep& 39\\
67 &\psep& 80 & 68 &\psep& 115 & 69 &\psep& 36 & 70 &\psep& 77 & 71 &\psep& 101 & 72 &\psep& 102\\
73 &\psep& 23 & 74 &\psep& 64 & 75 &\psep& 108 & 76 &\psep& 29 & 77 &\psep& 70 & 78 &\psep& 105\\
79 &\psep& 26 & 80 &\psep& 67 & 81 &\psep& 51 & 82 &\psep& 52 & 83 &\psep& 93 & 84 &\psep& 14\\
85 &\psep& 58 & 86 &\psep& 99 & 87 &\psep& 20 & 88 &\psep& 55 & 89 &\psep& 96 & 90 &\psep& 17\\
91 &\psep& 41 & 92 &\psep& 42 & 93 &\psep& 83 & 94 &\psep& 4 & 95 &\psep& 48 & 96 &\psep& 89\\
97 &\psep& 10 & 98 &\psep& 45 & 99 &\psep& 86 & 100 &\psep& 7 & 101 &\psep& 71 & 102 &\psep& 72\\
103 &\psep& 113 & 104 &\psep& 34 & 105 &\psep& 78 & 106 &\psep& 119 & 107 &\psep& 40 & 108 &\psep& 75\\
109 &\psep& 116 & 110 &\psep& 37 & 111 &\psep& 61 & 112 &\psep& 62 & 113 &\psep& 103 & 114 &\psep& 24\\
115 &\psep& 68 & 116 &\psep& 109 & 117 &\psep& 30 & 118 &\psep& 65 & 119 &\psep& 106 & 120 &\psep& 27\\
\hline
\end{array} 
\)
}
\end{center}
\end{table}

\begin{table}[htb]
\caption{Pairs of indexes for $J_2$-opposite states.}
\label{TabOppJ2}
\begin{center}
{\setlength{\arraycolsep}{0.15em}
\newcommand{\psep}{\leftrightarrow}
\(
\begin{array}{|rcl|rcl|rcl|rcl|rcl|rcl|}
\hline
1 &\psep& 21 & 2 &\psep& 22 & 3 &\psep& 24 & 4 &\psep& 23 & 5 &\psep& 28 & 6 &\psep& 30\\
7 &\psep& 29 & 8 &\psep& 25 & 9 &\psep& 27 & 10 &\psep& 26 & 11 &\psep& 31 & 12 &\psep& 32\\
13 &\psep& 34 & 14 &\psep& 33 & 15 &\psep& 38 & 16 &\psep& 40 & 17 &\psep& 39 & 18 &\psep& 35\\
19 &\psep& 37 & 20 &\psep& 36 & 21 &\psep& 1 & 22 &\psep& 2 & 23 &\psep& 4 & 24 &\psep& 3\\
25 &\psep& 8 & 26 &\psep& 10 & 27 &\psep& 9 & 28 &\psep& 5 & 29 &\psep& 7 & 30 &\psep& 6\\
31 &\psep& 11 & 32 &\psep& 12 & 33 &\psep& 14 & 34 &\psep& 13 & 35 &\psep& 18 & 36 &\psep& 20\\
37 &\psep& 19 & 38 &\psep& 15 & 39 &\psep& 17 & 40 &\psep& 16 & 41 &\psep& 101 & 42 &\psep& 102\\
43 &\psep& 104 & 44 &\psep& 103 & 45 &\psep& 108 & 46 &\psep& 110 & 47 &\psep& 109 & 48 &\psep& 105\\
49 &\psep& 107 & 50 &\psep& 106 & 51 &\psep& 111 & 52 &\psep& 112 & 53 &\psep& 114 & 54 &\psep& 113\\
55 &\psep& 118 & 56 &\psep& 120 & 57 &\psep& 119 & 58 &\psep& 115 & 59 &\psep& 117 & 60 &\psep& 116\\
61 &\psep& 81 & 62 &\psep& 82 & 63 &\psep& 84 & 64 &\psep& 83 & 65 &\psep& 88 & 66 &\psep& 90\\
67 &\psep& 89 & 68 &\psep& 85 & 69 &\psep& 87 & 70 &\psep& 86 & 71 &\psep& 91 & 72 &\psep& 92\\
73 &\psep& 94 & 74 &\psep& 93 & 75 &\psep& 98 & 76 &\psep& 100 & 77 &\psep& 99 & 78 &\psep& 95\\
79 &\psep& 97 & 80 &\psep& 96 & 81 &\psep& 61 & 82 &\psep& 62 & 83 &\psep& 64 & 84 &\psep& 63\\
85 &\psep& 68 & 86 &\psep& 70 & 87 &\psep& 69 & 88 &\psep& 65 & 89 &\psep& 67 & 90 &\psep& 66\\
91 &\psep& 71 & 92 &\psep& 72 & 93 &\psep& 74 & 94 &\psep& 73 & 95 &\psep& 78 & 96 &\psep& 80\\
97 &\psep& 79 & 98 &\psep& 75 & 99 &\psep& 77 & 100 &\psep& 76 & 101 &\psep& 41 & 102 &\psep& 42\\
103 &\psep& 44 & 104 &\psep& 43 & 105 &\psep& 48 & 106 &\psep& 50 & 107 &\psep& 49 & 108 &\psep& 45\\
109 &\psep& 47 & 110 &\psep& 46 & 111 &\psep& 51 & 112 &\psep& 52 & 113 &\psep& 54 & 114 &\psep& 53\\
115 &\psep& 58 & 116 &\psep& 60 & 117 &\psep& 59 & 118 &\psep& 55 & 119 &\psep& 57 & 120 &\psep& 56\\
\hline
\end{array} 
\)
}
\end{center}
\end{table} 

The matrices and entangled states coincide with derived earlier for 
Witting configuration in Ref.~\cite{dod}
\begin{equation}\label{entW1}
\ket{\Omega_1}=\frac{1}{2}\bigl(\ket{0}\ket{2}-\ket{2}\ket{0}+\ket{1}\ket{3}-\ket{3}\ket{1}\bigr)
\end{equation}
and
\begin{equation}\label{entW2}
\ket{\Omega_2}=\frac{1}{2}\bigl(\ket{0}\ket{1}-\ket{1}\ket{0}+\ket{3}\ket{2}-\ket{2}\ket{3}\bigr).
\end{equation}
The indexes for $J_1$- and $J_2$-opposite states are listed in Table~\ref{TabOppJ1} 
and Table~\ref{TabOppJ2} respectively.

\section{Conclusion and discussion}
\label{Sec:Concl}

New configuration with 120 quantum states useful for analysis of principles
of quantum mechanics such as nonlocality and contextuality was introduced in presented
work. Many properties of the configuration has analogies with models
based on Witting configuration \cite{WA,dod} and geometry of
dodecahedron \cite{WA,dod,ZP,shadows} already known earlier. 

{
\noindent Let us compare some properties of both complex configurations:\\
\begin{tabular}{|l||c|c|}\hline
            & Witting configuration & $\LSv$ \\ \hline \hline
 Dimension  & 4   & 4    \\ \hline        
 States     & 40  & 120  \\ 
 \ orthogonal to a state
            & 12  & 21   \\ \hline
 Bases      & 40  & 210  \\ 
 \ with a state    & 4  & 7  \\ \hline
 N-cliques  &  2970     & 31\,475\,754 \\
 \ minimal size & 4       & 10 \\
 \ maximal size & 7       & 24 \\
 \ required     & 10      & 30 \\ \hline
 Symmetry group  & $U_4(\mathbb{F}_2)$ & $A_7$ \\ 
 Size of the group   & 25920 & 2520 \\ \hline
 Equal bases & 10 & 30 \\ 
 Size of a subgroup&  1440 &  240 \\ \hline
 $\bigl|\brkt{\psi_j}{\psi_k}\bigr|^2\!,$ $j \neq k$ & $\vphantom{\Bigg|}$
  0, $\dfrac{1}{3}$& 
  0, $\dfrac{1}{7}$, $\dfrac{2}{7}$, $\dfrac{4}{7}$ \\ \hline
Polynomial & $\vphantom{\sum^2} z^2+z+1=0$ & $z^2+z+2=0$ \\ \hline  
\end{tabular}
\samepage

}

\medskip

Here two roots of {\em polynomial} may be used for construction of coordinates in the 
configurations, {\em N-cliques} are {\em maximal cliques with nonorthogonal states} used 
in brute-force search for (impossible) classical noncontextual models 
and {\em equal bases} respected by an appropriate subgroup suggest possibility 
to apply for entangled configurations the same transformation to both systems  
in a simpler measurement scheme. 

\medskip

It is also worth mentioning some related topics, the discussion of which lies beyond 
the purpose of this work. Model with two entangled quantum systems based on Witting polytope
can be used for description of quantum communications scheme \cite{WitQCom}, and 
the question regarding the potential for a similar use of configuration $\LSv$ might 
be quite reasonable.

In addition to this, there exists yet another configuration of quantum states based on
the complexification of $E_8$ discussed recently in Ref.~\cite{Ohta}. 
In a certain sense, this configuration is even more
directly related to $E_8$, because it is constructed by complexification of one of the
standard representations without orthogonal transformations in the 8D real space
that characterizes the both configurations discussed earlier.

Up to phase factor, there are only 60 {\em different} quantum states 
corresponding to the 240 complex vectors of this 
configuration.  They coincide with {\em stabilizer}
states for two qubits \cite{Ohta}.
Issues of contextuality related to the 105 bases that can be constructed for
these states were discussed in Ref.~\cite{WA60}.
However, in this case, the symmetry-based approach used for the other two 
configurations may require addressing a number of nontrivial issues \cite{dod25}.

\end{document}